\documentclass[11pt, notitlepage, letterpaper]{article}
\pdfoutput = 1
\usepackage[utf8]{inputenc}
\usepackage{amsfonts, amsmath, mathtools, enumerate, amssymb, hyperref, graphics, enumerate, dsfont, amsthm, csquotes, paralist, bbm, url}
\usepackage[english]{babel}
\usepackage[nice]{nicefrac}
\usepackage[margin=1in]{geometry}
\allowdisplaybreaks

\usepackage[sorting=ynt, url = false, doi = false, isbn =false, backend = biber]{biblatex}
\newtheorem{theorem}{Theorem}
\newtheorem{defn}{Definition}
\newtheorem{lemma}{Lemma}
\newtheorem{claim}{Claim}

\newcommand{\cA}{\mathcal{A}}

\newcommand{\cD}{\mathcal{D}}

\def\b0{{\bf 0}}
\def\b1{{\bf 1}}

\usepackage{textcomp}

\title{Secretary Problems: The Power of a Single Sample}
\author{Pranav Nuti \and Jan Vondr\'ak}
\date{\today}

\begin{document}
\maketitle

\begin{abstract}
In this paper, we investigate two variants of the secretary problem. In these variants, we are presented with a sequence of numbers $X_i$ that come from distributions $\mathcal{D}_i$, and that arrive in either random or adversarial order. We do not know what the distributions are, but we have access to a single sample $Y_i$ from each distribution $\mathcal{D}_i$. After observing each number, we have to make an irrevocable decision about whether we would like to accept it or not with the goal of maximizing the probability of selecting the largest number.

The random order version of this problem was first studied by Correa et al. [SODA 2020] who managed to construct an algorithm that achieves a probability of 0.4529. In this paper, we improve this probability to 0.5009, almost matching an upper bound of $\simeq 0.5024$ which we show follows from earlier work. We also show that there is an algorithm which achieves the probability of $\simeq 0.5024$ asymptotically if no particular distribution is especially likely to yield the largest number. For the adversarial order version of the problem, we show that we can select the maximum number with a probability of $1/4$, and that this is best possible. Our work demonstrates that unlike in the case of the expected value objective studied by Rubinstein et al. [ITCS 2020], knowledge of a single sample is not enough to recover the factor of success guaranteed by full knowledge of the distribution.
\end{abstract}

\thispagestyle{empty}
\newpage
\setcounter{page}{1}

\section{Introduction}

In the secretary problem, we observe a sequence of values (chosen adversarially) in a uniformly random order, with the goal of stopping at the maximum value with probability as large as possible. It is well known that this can be done with a probability of $\frac{1}{e}$. Over the years, many variants of the secretary problem have been studied. In this paper, we are particularly interested in variants where some additional information is available about the values before they arrive.

In the first variant, the numbers come from independent distributions $\mathcal{D}_i$ known to us, and arrive in a random order. This problem was introduced by Esfandiairi et al. \cite{Superstars}, and the work of Nuti \cite{me1} showed that it is possible to select the largest number with a probability of $\simeq 0.5801$, and that this is the best possible constant. In the second variant, also studied in \cite{Superstars}, in addition to coming from distributions, the observations are adversarially ordered. This variant is similar to a prophet inequality, but with the expected value objective replaced by the ``best-choice" objective. The best possible constant for this variant is $\frac{1}{e}$, as demonstrated in \cite{Superstars}.

The work of Rubinstein et al. \cite{rubinstein} revealed that for the purposes of obtaining a prophet inequality, where values from distributions $\cD_i$ arrive in adversarial order and the goal is to maximize the {\em expectation} of the selected value, knowledge of a single sample from each distribution can replace the knowledge of the entire distributions: One can obtain $\frac{1}{2}$ of the optimum in expectation and this is optimal in both models. The knowledge of a single prior sample rather than the whole distribution has been also studied for other variants of prophet inequalities, such as prophet inequalities under a random order, prophet inequalities under more complex constraints, and pricing problems which are often the motivating setting for these questions \cite{DRY15,sampledriven,DFLLR21a,DFLLR21b,CDFFLLP22}. 

Exactly how useful is access to a single prior sample for variants of the secretary problem? Our work is an attempt to answer this question.
%In the past, most work attempting to understand the power of a single sample has focused on the expected value objective (see for example the work of Correa et al. \cite{sampledriven} studying random order prophet inequalities, or the work of Caramanis et al. \cite{CDFFLLP22} extending the techniques of Rubinstein et al. \cite{rubinstein} to various combinatorial prophet inequalities). However, we wish to understand whether a single sample remains powerful for other kinds of objectives.
In particular, we study the following problem:
\begin{itemize}
    \item {\em The random (or adversarial) order single sample secretary problem:} Independent random variables $X_1,X_2,\ldots,X_n$ are revealed one by one in random (or adversarial) order. We do not know the distribution but we have a prior sample $Y_i$ from the same distribution as $X_i$, for each $i$. We can accept each revealed value or move on; the goal is to maximize the probability of accepting the largest value amongst the $X_i$. 
\end{itemize}
An important tool to study this question is the following closely related problem:
\begin{itemize}
    \item  {\em The random (or adversarial) order  two-sided Game of Googol:} $n$ cards with numbers on both sides are placed on a table in random (or adversarial)  order with each side facing up with probability $\frac{1}{2}$. We see all the face-up numbers. We can then flip  cards one by one, and upon seeing the revealed face-down number, accept the card or move on. The goal is to maximize the probability of accepting the highest (initially) face-down number.
\end{itemize}
The (random order) two-sided Game of Googol was first introduced by Correa et. al. \cite{googol}, where they managed to establish a lower bound of $\simeq 0.4529$ on the probability of success. Clearly, by letting the numbers on the two sides of a card be $X_i$ and $Y_i$, the random (resp. adversarial) order single sample secretary problem reduces to the random (resp. adversarial) order two-sided Game of Googol, so throughout this paper, we will be interested in both these problems.

\subsection{Our contributions}

In this paper, we obtain closely matching positive and negative results for the random and adversarial order single sample secretary problem. 

For the random order version of the problem, we are able to show:

\begin{theorem}
\label{thm:RO}
There is an algorithm that achieves a probability of success of $0.5009$ for the two-sided Game of Googol (and hence also for the single sample secretary problem). The algorithm uses only pairwise comparisons between values. 

Assuming that only pairwise comparisons are used, there is no algorithm for the single sample secretary problem (and hence also for the two-sided Game of Googol) achieving a probability of success better than $\gamma \simeq 0.5024$.
\end{theorem}

Our main contribution is the positive result, the first part of the theorem. The algorithm we consider uses the (initially) face-up numbers to compute a threshold for each face-down number that it will need to cross in order to be accepted. The analysis of the algorithm consists in being able to understand the quantity $\Pr[X^i > Y^j]$ well, where $X^i$ is the rank $i$ element amongst the face-down numbers, and $Y^j$ is the rank $j$ element amongst the face-up numbers. 

Especially crucial is a novel lemma about this quantity that might be of independent interest (we phrase it here informally, working with the case where the numbers on the two sides of the $i^\text{th}$ card are samples from a distribution $\mathcal{D}_i$): If we have two sets of samples from independent distributions $\mathcal{D}_1, \mathcal{D}_2, \ldots, \mathcal{D}_n$, the probability of a lower ranked element from one set beating a higher ranked element from the other set is maximized when the distributions are identical.

The hardness of $\gamma \simeq 0.5024$ follows from an optimality result discussed first in the work of Campbell and Samuels \cite{CS81} and then again in \cite{Independentsampling, advice}.  

It is natural to ask whether $\gamma \simeq 0.5024$ is in fact the optimal probability of success. We prove that this factor can be achieved for the single sample secretary problem under an additional ``no-superstars'' assumption (which has also been considered in recent related work  \cite{Superstars}).

\begin{defn}
\label{def:no-stars}
A collection of random variables $X_1,\ldots,X_n$ satisfies the no-$\epsilon$-superstars assumption, if none of the variables $X_i$ is the maximum among $X_1,\ldots,X_n$ with probability more than $\epsilon$.
\end{defn}

\begin{theorem}
\label{thm:no-stars}
There is an algorithm for the single sample secretary problem which achieves a factor of $\gamma - O(1 / \log \frac{1}{\epsilon})$ under the no-$\epsilon$-superstars assumption.
\end{theorem}

The main idea behind our proof of this theorem is intuitive: Given the no-$\epsilon$-superstars assumption, it is quite unlikely that some two of the largest $k=O(\log \frac{1}{\epsilon})$ numbers amongst all the $X_i$ and $Y_i$ come from the same distribution $\mathcal{D}_j$. This allows us to argue that the problem is quite close to the single sample secretary problem with identical distributions, which is known to have an optimal probability of success $\gamma \simeq 0.5024$.

We believe that the factor of $\gamma$ should be the optimal answer for the two-sided Game of Googol, as well as the single sample secretary problem; we leave this as an open problem. Note that in this paper, we have refuted the possibility that the answer is $\frac{1}{2}$, a conjecture one might perhaps make upon superficial examination of the problem, or after running simulations.

Next, we turn to the adversarial order version of the problem. We prove:

\begin{theorem}
\label{thm:AO}
There is an algorithm that achieves a factor of $\frac{1}{4}$ for the adversarial order two-sided Game of Googol (and hence also for the adversarial order single sample secretary problem).

There is no algorithm for the adversarial order single sample secretary problem (and hence also for the adversarial order two-sided Game of Googol) achieving a factor better than $\frac{1}{4}$.
\end{theorem}

Our main contribution here is the negative result, the second part of the theorem. The problem of choosing the maximum among the $X_i$'s is closely related to the problem of selecting the last occurring $1$ in a sequence of independent $0$'s and $1$'s. This problem was studied in \cite{Bruss2000} and solved by a theorem known as ``sum the odds to one and stop''. In order to prove the negative result, we study and solve a similar problem where instead of knowing the underlying distributions, we only have access to a single sample from each distribution. 

This requires an interesting application of Ramsey's theorem (also employed recently in a similar style in \cite{Correaramsey, obliviousOCRS}), and an analysis of an optimal stopping problem on a stochastic process using the tools of Ferguson \cite{Ferguson}.

The first part of the theorem, while not discussed in the literature previously, is  similar to recent results by Correa et al. \cite{Independentsampling}.  In their model, each of $n$ input values is independently selected with probability $\alpha$ to be a ``prior value'' which can only be observed. For $\alpha=1/2$, the problem has a flavor similar to ours. A natural algorithm for both models is to set the largest ``prior value" (our in our case, the largest (initially) face-up number) as a threshold, and accept the first number which beats this threshold. However, the techniques for the negative result in their model do \textit{not} seem to apply to our model directly. For the goal of understanding the power of a single sample, perhaps our model is more natural.

Our results indicate that in the context of the ``best-choice" objective of secretary problems, unlike for the prophet inequality, the knowledge of a single prior sample is strictly weaker than the knowledge of the whole distribution. 

\subsection{Organization}
In section 2, we discuss of the random order variant of the problem, and prove Theorem 1. In section 3, we discuss the no-$\epsilon$ superstars assumption, and prove theorem 2. Finally, in section 4, we study the adversarial order variant of the problem.

\section{The random order variant}

In this section we will prove theorem \ref{thm:RO}. We start by quickly indicating the proof of the negative result.

\paragraph{Negative result:} The hardness of $\gamma \simeq 0.5024$ follows from a known optimality result for the following problem:

\begin{itemize}
\item {\em The secretary problem with $\alpha n$ prior values}: An unknown sequence of values $x_1,\ldots,x_n$ arrives in a random order. The first $\alpha n$ arriving values can be only observed. Then we can accept any revealed value or move on; the goal is to choose the maximum of the last $(1-\alpha) n$ values. 
\end{itemize}

This problem, under the name of ``choosing the best of the current crop," was discussed and solved in a paper from 1981 \cite{CS81}, and similar results have been independently rediscovered in \cite{Independentsampling, advice}. The case most relevant to us is $\alpha=\frac12$, where the first half of the input can only be observed, and we want to pick the maximum from the second half. The optimal probability in this case was determined to be $\gamma \simeq 0.5024$, assuming that only pairwise comparisons between values are allowed. 

If the $n$ values $x_1,\ldots,x_n$ are realizations of $\frac{n}{2}$ independent identically distributed random variables, then we obtain a special case of the single sample secretary problem. Hence, the value $\gamma \simeq 0.5024$ is an upper bound for the optimal probability of success for the single sample secretary problem, provided that only pairwise comparisons are allowed. 

\paragraph{Positive result:} We will concentrate now on analysing the following algorithm for the two-sided Game of Googol\footnote{Let us note that as $n$ increases, the optimal probability of success only gets worse, since it is possible to embed the problem with $n$ cards in the problem with $n+1$ cards by adding a card that has $0$ on both sides. This means that we can work in the limit as $n \to \infty$, and for convenience, we will do this.}:

\paragraph{The block-rank algorithm.}
Fix constants $1 \geq c_0 \geq c_1 \geq \cdots \geq c_r \geq \cdots \geq 0$ (we will choose these later). Generate $n$ uniformly random \textit{times} in $[0, 1]$, and let us say that the $k^\text{th}$ card arrives at the $k^\text{th}$ \textit{largest} time. (Note that somewhat peculiarly, time runs from $1$ to $0$, but this simplifies some notation later). 

Call a number special if it is a face-down number, arrives at a time $\leq c_j$, has rank less than or equal to $j + 1$ amongst the face-up numbers, and is the largest number amongst the face-down numbers observed thus far (for any $j$). Pick the first special number.

\

In other words, if a face-down number arrives at a time between $c_j$ and $c_{j-1}$, in order to be accepted, it must not only be larger than all the face-down numbers that have been observed thus far, but must also beat the $j+1^\text{th}$ ranked face-up number. As the time gets closer to $0$, we become increasingly lax in our demands about how many face-up numbers you need to beat in order to be accepted. 

This algorithm has an extremely useful monotonicity property-- if at some point during the running of the algorithm it were to decide to accept a number, and we artificially make it reject, then it would also wish to accept the first number it sees bigger than the number we made it reject.

Due to this monotonicity, the last special number (if there is one) must be the largest number amongst the face-down numbers. Hence, our algorithm that picks the first special number picks the largest number (and succeeds) if and only if there is exactly one special number. 

Therefore, we now wish to compute the probability of there being only one special number. In order to do so, we require some notation:

Let $X^i$ be the rank $i$ element amongst the face-down numbers, and let $Y^j$ be the rank $j$ element amongst the face-up numbers (with the understanding that $X^0 = Y^0 = \infty$; note the non-standard use of the superscript, which does {\em not} denote an exponent.) Finally, let us introduce the function $$H_j(x) = \sum_{i = 0}^\infty \Pr[X^{i+1} > Y^{j+1}]x^i$$

\begin{lemma}
The probability of success of the block-rank algorithm is:
$$\sum_{j = 0}^{\infty}\int_{c_{j+1}}^{c_j}\left(H_j(0)-\int_0^y  \frac{H_j(x)-H_j(0)}{x}dx\right)dy$$
\end{lemma}

\begin{proof}

The probability that there is at least one special number is the same as the probability that there is a \textit{last} special number, which equals:
\begin{align*}
    &\sum_{j = 0}^{\infty} \Pr[\text{the last special number arrives at a time between } c_j \text{ and } c_{j-1}]\\
    &=\sum_{j = 0}^{\infty} \Pr[\text{the rank 1 face-down number arrives at a time between } c_j \text{ and } c_{j-1} \text{ and is special}]\\
    &= \sum_{j = 0}^{\infty}(c_j-c_{j-1})\Pr[X^1>Y^{j+1}]\\
    &=\sum_{j = 0}^{\infty}(c_j-c_{j+1})H_j(0)
\end{align*}

where the first equality follows from monotonicity. The probability that there are at least two special numbers is the same as the probability there is a \textit{second last} special number. Clearly, summing over the possibility that the second last special number has rank $i+1$ amongst the face-down numbers, this equals
\begin{align*}
    &\sum_{j = 0}^{\infty}\int_{c_{j+1}}^{c_j}\sum_{i = 1}^{\infty} \Pr[\text{the second last special number arrives at time between } y \text{ and } y+dy \text{ and has rank } i+1] 
\end{align*}

Note that for the rank $i+1$ number to be the second last special number, all the $i$ numbers bigger than it must arrive at lower times than it, and those $i$ numbers must be arranged so that the rank $1$ number comes first (once again, due to monotonicity). Therefore, the above is actually equal to:
$$\sum_{j = 0}^{\infty}\int_{c_{j+1}}^{c_j}\sum_{i = 1}^{\infty} \Pr[X^{i+1}>Y^{j+1}]\frac{y^i}{i}dy =\sum_{j = 0}^{\infty}\int_{c_{j+1}}^{c_j}\int_0^y  \frac{H_j(x)-H_j(0)}{x}dxdy$$
So the desired conclusion follows
\end{proof}

Our next goal is to establish a lower bound on this expression. Therefore, we wish to understand the coefficients of $H_j$ better. In order to do that, we introduce some further notation.  

Denote by $a_1 > a_2 > \cdots > a_{2n}$ the set of all numbers on the cards (including both face-up and face-down numbers). We say $a_m$ and $a_n$ are a \textit{pair} if they lie on the same card. Furthermore, let us use the shorthand $a \in X$ to say that $a$ is a face-down number, and $a \in Y$ to say that $a$ is a face-up number. Finally, let us define 

$$p(i+1, j+1) = \sum_{k=0}^j\frac{1}{2^{i+k+1}}\binom{i+k}{k}$$

The following facts explain our choice of definitions:

\paragraph{Fact 1.} Whether $X^{i+1} > Y^{j+1}$ depends only on which of $a_1,a_2,\ldots,a_{i+j+1}$ are in $X$ and which are in $Y$. 

\paragraph{Fact 2.} \label{fact2} Suppose there are no pairs amongst $a_1,a_2,\ldots,a_{i+j+1}$ (i.e., they all lie on different cards). Then,
 \begin{align*}
     \Pr[X^{i+1} > Y^{j+1}] &=\sum_{k=0}^j\Pr[Y^{k}>X^{i+1} > Y^{k+1}]\\
     &=\sum_{k=0}^j\frac{1}{2^{i+k+1}}\binom{i+k}{k}\\
     &= p(i+1, j+1)
 \end{align*}

Let us remark here on the connection between our approach and the optimal constant $\gamma\simeq 0.5024$. If the numbers on the cards were chosen to be i.i.d.~realizations of a distribution, in the limit as $n \to \infty$, we would have $ \Pr[X^{i+1} > Y^{j+1}] = p(i+1, j+1)$ since it would basically never happen that there are pairs amongst $a_1,a_2,\ldots,a_{i+j+1}$. If we analyze the probability of the block-rank algorithm's success in this case, and choose the optimal values for the $c_j$, we would obtain the optimal constant $\gamma\simeq 0.5024$. Thus, understanding the difference between $\Pr[X^{i+1} > Y^{j+1}] $ and $p(i+1, j+1)$ will let us understand how close we can get to $\gamma$.

\

The following lemma is therefore critical to our analysis:

\begin{lemma}
\label{lemma:critical}
If $i \leq j$, then $\Pr[X^i > Y^j] \geq p(i, j)$ (and if $i \geq j$, then $\Pr[X^i > Y^j] \leq p(i, j)$).
\end{lemma}
\begin{proof}
 Let us assume $i\leq j$. We will prove the statement by induction on the number of pairs amongst $a_1, a_2, \ldots, a_{i+j-1}$ (i.e., the number of pairs of numbers which lie on the same card). Clearly, the statement is true for $0$ pairs. So let us now assume the statement is true for $l-1$ pairs, and prove it for $l$ pairs. Let the lower numbers of the $l$ pairs be $a_{k_1}>a_{k_2}> \cdots > a_{k_l} $.
 
\begin{claim}
Let $k_0 = 0, k_{l+1} = \infty$. Then, for $0\leq s\leq l$, we have $$\Pr[Y^k>X^i> Y^{k+1}] = \begin{cases}
  \frac{1}{2^{i+k-2s}}\binom{i+k-2s-1}{k-s}  & \text{ if } k_{s+1}-i>k>k_s-i \\
  \frac{1}{2^{i+k-2s+1}}\binom{i+k-2s}{k-s}  & \text{ if } k = k_s-i
\end{cases}$$
\end{claim}
\begin{proof}
We will explain the claim for $k = k_s - i$, the proof for the other case is similar. For it to be true that $Y^k>X^i> Y^{k+1}$, we must have $a_{i+k} \in X$, and amongst the $i+k - 2s$ numbers amongst $a_1, a_2, \ldots, a_{i+k}$ that are not in pairs, we must select $k-s$ which will be in $Y$ (there are $\binom{i+k-2s}{k-s}$ ways to do this). Once we have done that, for $a_{i+k}$ and all numbers that are not in pairs, it is fixed whether they belong to $X$ or $Y$. Accordingly we must divide by $2^{i+k-2s+1}$.
\end{proof}

Returning to the proof of the lemma, note that $\Pr[X^i> Y^j] = \sum_{k = 0}^{j-1}\Pr[Y^k>X^i> Y^{k+1}]$. It suffices to show that this quantity decreases if the last pair, $k_l$, is not present. Now most of the terms in the summation are the same, whether the last pair is present or not. Keeping only the terms that are different, we need to show:
$$\frac{1}{2^{k_l-2l+1}}\binom{k_l-2l}{k_l-i-l} + \sum_{k =k_l-i+1}^{j-1}\frac{1}{2^{i+k-2l}}\binom{i+k-2l-1}{k-l}\geq \sum_{k =k_l-i}^{j-1}\frac{1}{2^{i+k-2(l-1)}}\binom{i+k-2(l-1)-1}{k-(l-1)}$$
Let us introduce the new constants $k_l - 2l  = \alpha$, $i - l = \beta, i+j-2l-1 = \gamma$ and a new dummy variable $t = i+k-2l$. Then the above is equivalent to:
\begin{alignat*}{2}
    &&\frac{1}{2^{\alpha+1}}\binom{\alpha}{\beta} + \sum_{t =\alpha+1}^{\gamma}\frac{1}{2^{t}}\binom{t-1}{\beta-1}
    &\geq \sum_{t =\alpha}^{\gamma}\frac{1}{2^{t+2}}\binom{t+1}{\beta}\\
    &\iff &\frac{2}{2^{\alpha}}\binom{\alpha}{\beta} + \sum_{t =\alpha+1}^{\gamma}\frac{4}{2^{t}}\binom{t-1}{\beta-1}&\geq \sum_{t =\alpha}^{\gamma}\frac{1}{2^{t}}\binom{t+1}{\beta}\\
    &\iff &\frac{2}{2^{\alpha}}\binom{\alpha}{\beta} + \sum_{t =\alpha}^{\gamma-1}\frac{2}{2^{t}}\binom{t}{\beta-1}&\geq \sum_{t =\alpha}^{\gamma}\frac{1}{2^{t}}\left(\binom{t}{\beta}+\binom{t}{\beta-1}\right)\\
    &\iff &\frac{2}{2^{\alpha+1}}\binom{\alpha+1}{\beta} + \sum_{t =\alpha+1}^{\gamma-1}\frac{2}{2^{t}}\binom{t}{\beta-1}&\geq \sum_{t =\alpha+1}^{\gamma}\frac{1}{2^{t}}\left(\binom{t}{\beta}+\binom{t}{\beta-1}\right)\\
    &&\cdot \\
    &&\cdot \\
    &&\cdot \\
    &\iff &\frac{2}{2^{\gamma-1}}\binom{\gamma-1}{\beta} + \frac{2}{2^{\gamma-1}}\binom{\gamma-1}{\beta-1}&\geq \sum_{t =\gamma-1}^{\gamma}\frac{1}{2^{t}}\left(\binom{t}{\beta}+\binom{t}{\beta-1}\right)\\
    &\iff &\frac{1}{2^{\gamma-1}}\binom{\gamma-1}{\beta} + \frac{1}{2^{\gamma-1}}\binom{\gamma-1}{\beta-1}&\geq \frac{1}{2^{\gamma}}\left(\binom{\gamma}{\beta}+\binom{\gamma}{\beta-1}\right)\\
    &\iff &2\binom{\gamma}{\beta}&\geq \binom{\gamma}{\beta}+\binom{\gamma}{\beta-1}\\
    &\iff &\binom{\gamma}{\beta}&\geq \binom{\gamma}{\beta-1}\\
    &\iff &\frac{\gamma+1}{2} &\geq \beta\\
    &\iff &\frac{i+j-2l}{2} &\geq i-l\\
    &\iff &j &\geq i
\end{alignat*}
So the desired result follows. The case $i \geq j$ is similar and follows from the same chain of implications above. Examining the proof closely, we see that we can actually divide it into two parts:
\begin{itemize}
    \item We can assume $k_l = i+j-1$ without changing the value of $\Pr[X^i > Y^j]$ (in fact this shows $\Pr[X^i > Y^j]$ depends only on the number of pairs, and not their positions).
    \item  $\Pr[Y^{j-1}> X^i> Y^j]$ is higher if $k_l = i+j-1$ than if $k_l = \infty$.
\end{itemize}\end{proof}

We have another lemma, further examining the coefficients of $H_j$, which is important to obtain the particular constant we want.

\begin{lemma}
$\Pr[X^2>Y^3]-p(2, 3)\leq 2(\Pr[X^1>Y^3]-p(1, 3))+(\Pr[X^1>Y^4]-p(1, 4))$.
\end{lemma}
\begin{proof} The proof consists of 3 cases:

\paragraph{Case 1:} There are no pairs amongst $a_1, a_2, a_3, a_4$. Then both sides of the expression are 0, so the result is true. 

\paragraph{Case 2:} There is exactly one pair amongst $a_1, a_2, a_3, a_4$. To calculate $\Pr[X^2>Y^3]$, we use the following formula:
\begin{align*}
    &\Pr[X^2>Y^3]\\ 
    &= \Pr[X^2>Y^2] + \Pr[Y^2>X^2>Y^{3}]\\
    &= \frac{1}{2}+\Pr[a_4, a_1 \in X, a_2, a_3 \in Y]+\Pr[a_4, a_2 \in X, a_1, a_3 \in Y]+\Pr[a_4, a_3 \in X, a_1, a_2 \in Y]
\end{align*}
We obtain that $\Pr[X^2>Y^3] = \frac{12}{16}$ (since if there is one pair, exactly two of the terms in the sum must be $\frac{1}{8}$, and the other term must be 0). It is easy to see $\Pr[X^1>Y^4] = 1$, and we know $p(2, 3)=\frac{11}{16}$, and $p(1, 4)=\frac{15}{16}$ from our earlier formula for $p(i+1, j+1)$, so the result is true, even with just the second term on the right hand side. 

\paragraph{Case 3:}There are two pairs amongst $a_1, a_2, a_3, a_4$. Then, it is easy to see $\Pr[X^2>Y^3] = 1$, $\Pr[X^1>Y^3] = 1, \Pr[X^1>Y^4] = 1$, and we can compute $p(2, 3)=\frac{11}{16}$, $p(1, 3)=\frac{7}{8},p(1, 4)=\frac{15}{16}$, so the result is true.
\end{proof}

We are finally prepared to analyze the expression of the algorithm's probability of success which, recall, equals:

$$\sum_{j = 0}^{\infty}\int_{c_{j+1}}^{c_j}\left(H_j(0)-\int_0^y  \frac{H_j(x)-H_j(0)}{x}dx\right)dy$$

For the terms $j \geq 3$, let us use the simple upper bound $\Pr[X^1 > Y^{j+1}] \geq \Pr[X^{i+1} > Y^{j+1}]$ to get $H_j(x) \leq H_j(0) \sum_{i = 0}^{\infty}x^i  = \frac{H_j(0)}{1-x}$. Setting $c_0 = 1, c_1 = 0.715598, c_2 = 0.496376, c_3 = c_4 = \cdots = c_N= 0.301284, c_{N+1} = 0$ for very large $N$, it follows that--
\begin{align*}
    &\sum_{j = 3}^{\infty}\int_{c_{j+1}}^{c_j}\left(H_j(0)-\int_0^y  \frac{H_j(x)-H_j(0)}{x}dx\right)dy\\ &\geq \sum_{j = 3}^{\infty}\int_{c_{j+1}}^{c_j}\left(H_j(0)-\int_0^y  \frac{\frac{H_j(0)}{1-x}-H_j(0)}{x}dx\right)dy\\
    &= \sum_{j = 3}^{\infty}\int_{c_{j+1}}^{c_j}\left(H_j(0)-H_j(0)\int_0^y  \frac{1}{1-x}dx\right)dy\\
    &\geq \sum_{j = 3}^{\infty}p(1, j)\int_{c_{j+1}}^{c_j}(1+ \log(1-y))dy\\
    &= \sum_{j = 3}^{\infty}\left(1-\frac{1}{2^{j+1}}\right)\int_{c_{j+1}}^{c_j}(1+ \log(1-y))dy\\
    &= \int_{0}^{c_3}(1+ \log(1-y))dy-  \sum_{j = 3}^{\infty}\frac{1}{2^{j+1}}\int_{c_{j+1}}^{c_j}(1+ \log(1-y))dy\\
    &= -(1-c_3)\log (1-c_3)-  \sum_{j = 3}^{\infty}\frac{1}{2^{j+1}}\int_{c_{j+1}}^{c_j}(1+ \log(1-y))dy\\
    &\geq 0.2504
\end{align*}
Recall lemma 1 which showed that $\Pr[X^i > Y^j] \leq p(i, j)$ for $i \geq j$, and note that a simple calculation shows $\sum_{i = 0}^\infty p(i+1, j+1)x^i = \sum_{k = 0}^j(2-x)^{-k-1}$. Therefore, for $j = 0$ we have
\begin{align*}
    \int_{c_{1}}^{c_0}\left(H_0(0)-\int_0^y  \frac{H_0(x)-H_0(0)}{x}dx\right)dy &\geq \int_{c_{1}}^{c_0}\left(\frac{1}{2}-\int_0^y  \frac{\frac{1}{2-x}-\frac{1}{2}}{x}dx\right)dy\\
    &-\int_{c_{1}}^{c_0}\frac{y^3}{3}(\Pr[X^4>Y^1]-p(4, 1))dy\\
    &\geq 0.0621 + (\Pr[X^1>Y^4]-p(1, 4))\int_{c_{1}}^{c_0}\frac{y^3}{3}dy
\end{align*}
where we have used the definition of $H$ and applied lemma 1 for most terms in the series, but carefully preserved one of the terms which will be useful later for the analysis. Similarly for $j=1$ we have
\begin{align*}
    \int_{c_{2}}^{c_1}\left(H_1(0)-\int_0^y  \frac{H_1(x)-H_1(0)}{x}dx\right)dy  &\geq \int_{c_{2}}^{c_1}\left(\frac{3}{4}-\int_0^y  \frac{\frac{1}{2-x}+\frac{1}{(2-x)^2}-\frac{3}{4}}{x}dx\right)dy\\
    &-\int_{c_{2}}^{c_1}\frac{y^2}{2}(\Pr[X^3>Y^2]-p(3, 2))dy\\
    &\geq 0.0809 + (\Pr[X^2>Y^3]-p(2, 3))\int_{c_{2}}^{c_1}\frac{y^2}{2}dy
\end{align*}
Finally, for $j = 2$, we have:
\begin{align*}
    &\int_{c_{3}}^{c_2}\left(H_2(0)-\int_0^y  \frac{H_2(x)-H_2(0)}{x}dx\right)dy \\ &\geq \int_{c_{3}}^{c_2}\left(\frac{7}{8}-\int_0^y  \frac{\frac{1}{2-x}+\frac{1}{(2-x)^2}+\frac{1}{(2-x)^3}-\frac{7}{8}}{x}dx\right)dy- \int_{c_{3}}^{c_2}y(\Pr[X^2>Y^3]-p(2, 3))dy\\
    &+\int_{c_{3}}^{c_2}(\Pr[X^1>Y^3]-p(1, 3))dy\\
    &\geq 0.1073 + \int_{c_{3}}^{c_2}(\Pr[X^1>Y^3]-p(1, 3)) - y(\Pr[X^2>Y^3]-p(2, 3))dy\\
\end{align*}
To obtain the desired result, we just need to observe:
\begin{align*}
   &(\Pr[X^2>Y^3]-p(2, 3))\left(\int_{c_{3}}^{c_2}y dy - \int_{c_{2}}^{c_1}\frac{y^2}{2}dy\right)\\
   &\leq  0.0372(\Pr[X^2>Y^3]-p(2, 3)) \\
   &\leq 0.1950(\Pr[X^1>Y^3]-p(1, 3))+ 0.0614(\Pr[X^1>Y^4]-p(1, 4))\\
   &\leq(\Pr[X^1>Y^3]-p(1, 3))\int_{c_{3}}^{c_2}dy+(\Pr[X^1>Y^4]-p(1, 4))\int_{c_{1}}^{c_0}\frac{y^3}{3} dy \\
\end{align*}
where the inequality in the middle follows from lemma 2. To conclude, we just add up all the terms above to see that this algorithm achieves a success probability of at least 0.5007 (and in fact, at least 0.5009, as we can observe by rounding to 5 decimal places instead of 4 in the calculation).

\section{The no-superstars assumption}

In this section, we prove that the optimal factor of $\gamma \simeq 0.5024$ can be achieved asymptotically when we assume that no particular element contributes significantly to the optimum. 

Recall Definition~\ref{def:no-stars}: We say that $X_1,\ldots,X_n$ satisfy the no-$\epsilon$-superstars assumption, if for every $i \in [n]$, $\Pr[X_i = \max_{1 \leq j \leq n} X_j] \leq \epsilon$.

As before, $Y_1,\ldots,Y_n$ denote the samples from the respective distributions, which are available to us. $X_1,\ldots,X_n$ are the numbers revealed in a random order we must accept or reject. 

We prove the following lemma.

\begin{lemma}
Under the no-$\epsilon$-superstars assumption, it happens with probability at most $\epsilon k 2^{k-1}$ that there is a pair $(X_i,Y_i)$ such that both $X_i$ and $Y_i$ are among the top $k$ values overall.
\end{lemma}

\begin{proof}
Suppose that we first generate the set of all $X_i$ and $Y_i$ values and then we decide independently for each pair which of the two values is $X_i$ and which is $Y_i$. We can assume without loss of generality (by some fixed tie-breaking rule) that all the values are distinct. 

Let us denote by $E_i$ the event that $(X_i,Y_i)$ is the pair maximizing $\min \{X_i,Y_i\}$, both $X_i, Y_i$ are among the top $k$ values overall, and $X_i < Y_i$. Let us also denote by $E'_i$ a sub-event of $E_i$, where in addition we require that all the values above $X_i$ are $Y$-values. Observe that conditioned on $E_i$ and the complete set of values (but not which of the other values are $X_{i'}$ or  $Y_{i'}$), it happens with probability at least $1/2^{k-2}$ that all the values above $X_i$ are $Y$-values. This is because the number of such values is at most $k-1$, and each of them (except $Y_i$) is independently chosen to be $X_{i'}$ or $Y_{i'}$. Hence,
$$ \Pr[E'_i] \geq \frac{1}{2^{k-2}} \Pr[E_i].$$

Next, we analyze the probability of $E'_i$. This event can only happen if $Y_i$ is among the top $k$ $Y$-values, and $X_i$ is the top $X$-value.
We have 
\begin{align*}
    \Pr[E'_i] & \leq \Pr[Y_i \text{ is among the top } k \ Y\text{-values}, \text{ and } X_i \text{ is the maximum } X\text{-value}] \\
    &=  \Pr[Y_i \text{ is among the top } k \ Y\text{-values}] \Pr[ X_i \text{ is the maximum } X\text{-value}] \\
& \leq  \epsilon \, \Pr[Y_i \text{ is among the top } k \ Y\text{-values}].
\end{align*}
Let us sum up over the indices $i$:
\begin{align*}
\sum_{i=1}^{n} \Pr[E'_i]   & \leq \sum_{i=1}^{n} \Pr[Y_i \text{ is among the top } k \ Y\text{-values}, \text{ and } X_i \text{ is the maximum } X\text{-value}] \\
& \leq \epsilon \sum_{i=1}^{n}  \Pr[Y_i \text{ is among the top } k \ Y\text{-values}] = \epsilon k
\end{align*}
because $\sum_{i=1}^{n} \Pr[Y_i \text{ is among the top } k \ Y\text{-values}]$ is simply the expected number of $Y$-values among the top $k$, which is exactly $k$.
Finally, as we argued above,
$$ \sum_{i=1}^{n} \Pr[E_i] \leq 2^{k-2}  \sum_{i=1}^{n} \Pr[E'_i] \leq \epsilon k 2^{k-2}.$$
By symmetry, $\sum_{i=1}^{n} \Pr[E_i]$ is exactly half of the probability that we aim to analyze --- the cases where the top pair $(X_i,Y_i)$ is among the top $k$ values and it is ordered so that $X_i < Y_i$. Hence, the probability that there is a pair $(X_i,Y_i)$ with both values among the top $k$ overall is at most $\epsilon k 2^{k-1}$. 
\end{proof}

The result follows now from the following lemma.

\begin{lemma}
\label{lemma:k-block}
Consider an instance of the two-sided game of Googol where no pair of numbers among the top $2k+1$ appears on the same card. Then the block-rank algorithm achieves a probability of success $\gamma - O(1/k)$ (where $\gamma \simeq 0.5024$ is the upper bound on the performance of any algorithm in the comparison model).
\end{lemma}

\begin{proof}
Let us return to the setup from the previous section, and suppose that amongst $a_1, a_2, \cdots, a_{2k+1}$, no two lie on the same card. Suppose $j \leq k$. Then:

\paragraph{Fact 3.} $\Pr[X^1> Y^{j+1}] = p(1, j+1)$.\\ This is a direct consequence of Fact \ref{fact2}.

\paragraph{Fact 4.} $\Pr[X^{i+1}> Y^{j+1}] \leq p(i+1, j+1)$ for every $i \geq 1$. \\
This is true because either $i \geq j$, in which case it follows from Lemma \ref{lemma:critical}, or else $i + j \leq 2k$, in which case it follows from Fact \ref{fact2}.

\

Hence, it follows that if we define $F_j(x) = \sum_{i = 0}^\infty p(i+1, j+1)x^i = \sum_{k = 0}^j(2-x)^{-k-1}$, we must have

$$\sum_{j = 0}^{k}\int_{c_{j+1}}^{c_j}\left(H_j(0)-\int_0^y  \frac{H_j(x)-H_j(0)}{x}dx\right)dy \geq \sum_{j = 0}^{k}\int_{c_{j+1}}^{c_j}\left(F_j(0)-\int_0^y  \frac{F_j(x)-F_j(0)}{x}dx\right)dy$$

We can interpret both sides of the above the expression as the success probability of a block-rank algorithm with $c_{k+1} = c_{k+2} = \ldots  = 0$. Now from earlier work in the literature analyzing the case of the cards having numbers that are i.i.d. samples\footnote{The notation is quite different from ours,  but this follows from lemma 3 and lemma 5 in \cite{Independentsampling}, for example.} (or the ``best of the current crop" setting) we know that if we choose the constants $c_j$ optimally, then 
$$\sum_{j = 0}^{\infty}\int_{c_{j+1}}^{c_j}\left(F_j(0)-\int_0^y  \frac{F_j(x)-F_j(0)}{x}dx\right)dy = \gamma.$$
Furthermore, it must be the case that 
$$\sum_{j = k+1}^{\infty}\int_{c_{j+1}}^{c_j}\left(F_j(0)-\int_0^y  \frac{F_j(x)-F_j(0)}{x}dx\right)dy = O(c_{k})$$
since this is the difference between the performance of two algorithms which differ in their behavior only in the last $c_{k}$ fraction of the face-down numbers. Finally once again from previous results in the literature\footnote{Similarly, this follows from lemma 4 in \cite{Independentsampling}.}, the optimal choice of $c_j$ satisfies $\int_{0}^{c_j}\frac{(1-x)^{-j-1}-1}{x}dx = 1$, and $c_k = O(1/k)$, so the desired result follows.
\end{proof}

Hence, we can complete the proof of Theorem~\ref{thm:no-stars} as follows.
Let us ignore the contributions from the cases where a card with both sides among the top $k$ values exists. Since this happens with probability at most $\epsilon k 2^{k-1}$, we obtain a factor at least $(1 - \epsilon k 2^{k-1}) (\gamma - O(1/k))$. This argument holds for any $k$; we can choose $k = \frac12 \log \frac{1}{\epsilon}$ and obtain $(1 - \frac12 \sqrt{\epsilon} \log \frac{1}{\epsilon})(\gamma - O(1/\log \frac{1}{\epsilon})) = \gamma - O(1/\log \frac{1}{\epsilon})$, using the fact that $\sqrt{\epsilon} = O(1 / \log^2 \frac{1}{\epsilon})$.

\section{The adversarial order variant}

In this section, we wish to establish Theorem \ref{thm:AO}. We start by establishing the positive result by examining the adversarial order two-sided Game of Googol.

\paragraph{Positive result.}
Our algorithm is simple: Set the largest (initially) face-up number as a threshold, and accept the first number which beats this threshold. The proof that this algorithm achieves a probability of $\frac{1}{4}$ is also simple: Suppose the two largest numbers that appear on the cards are $a_1$ and $a_2$. Note that the algorithm succeeds as long as $a_2$ is face-up and $a_1$ is face-down. Now there are two possibilities:

\textbf{Case 1:} $a_1$ and $a_2$ lie on different cards. In this case, $a_2$ is face-up and $a_1$ is face-down with probability $\frac{1}{4}$.

\textbf{Case 2:} $a_1$ and $a_2$ lie on the same card. In this case, $a_2$ is face-up and $a_1$ is face-down with probability $\frac{1}{2}$.

So in either case we win a probability of at least $\frac{1}{4}$. 

\paragraph{Negative result.}
To begin our analysis of the negative result, let us consider the following distribution:
$$X_i \text{ and } Y_i = \begin{cases}
     i & \text{with probability } p_i\\
     0  &  \text{with probability } 1-p_i\\
    \end{cases}$$
It is clear for this choice of distributions that the problem of picking the largest $X_i$ is the same as picking the last nonzero $X_i$.

Therefore, in order to prove the negative result, it suffices to show that it is impossible to solve the following problem with a probability strictly better than $\frac{1}{4}$:
\begin{itemize}
    \item {\em The single-sample last success problem:} Independent Bernoulli random variables $X_1,X_2, $ $\ldots,X_N$ are revealed one by one. We do not know the distribution, but we have a prior sample $Y_i$ from the same distribution as $X_i$, for each $i$. We can accept each revealed value or move on; the goal is to maximize the probability of accepting the last 1 amongst the $X_i$. 
\end{itemize}

To begin proving this result we first show that it is impossible to do strictly better than $\frac{1}{4}$ for a particular distribution of inputs and a particular class of stopping rules\footnote{Let us note that we are not trying to show this result for a fixed $N$ (in fact, it would be false for a fixed $N$). What we are trying to show is that in the limit as $N \to \infty$ there is no algorithm which does strictly better than $\frac{1}{4}$}. We will then leverage this to prove the general result with an application of Ramsey's theorem. 

As above, let us consider $X_1, X_2, \ldots, X_n$ (and $Y_1, Y_2, \ldots, Y_n$), independent identically distributed Bernoulli random variables which are 1 with probability $\frac{1}{n^{2/3}}$. In the following, we will condition on there not existing a $j$ for which $X_j = Y_j = 1$. Since this is an event with a probability equal to 1 in the limit as $n \to \infty$, it will not affect our results.

Suppose that the $k^{\text{th}}$ 1 amongst the $X_i$ occurs at $X_{t(k)}$. Let us define a random variable $A_k$ to be the total number of 1s that occur amongst  $X_1, X_2, \ldots, X_{t(k)}$ \textbf{and} $Y_1, Y_2, 
\ldots, Y_{t(k)-1}$ (note that $Y_{t(k)} \neq 1$ since $X_{t(k)} = 1$). Similarly, we define $B_k$ to be the number of 1s that amongst the $Y_{t(k)+1},\ldots Y_n$. 
We would like to establish that when we observe the $k^{\text{th}}$ 1 amongst the $X_i$, and we have to decide whether we would like accept it or not while knowing only $A_k$ and $B_k$, it is impossible to accept the last 1 with a probability strictly better than $\frac{1}{4}$. We formalize this as follows.

We define a stochastic process $(Z_k)_{k=1}^{\infty}$, where
$Z_k = (A_k, B_k)$ for $1 \leq k \leq K = \sum_{i=1}^{n} X_i$, and $Z_k = \varnothing$ for $k > K$. We consider a new problem where the $Z_k$'s are revealed one by one and the goal is to stop at $Z_K$, i.e. the last time before $Z_k$ becomes $\varnothing$.

\begin{lemma}
There is no stopping rule adapted to $\mathcal{G}_k = \sigma(Z_1, Z_2, \ldots, Z_k)$ (i.e., a stopping rule that makes decisions using only the values of $Z_1, Z_2, \ldots, Z_k$ at time $k$)  which succeeds with a probability $\geq \frac{1}{4}+\epsilon$ (for $\epsilon>0$ independent of $n$) at stopping at time $k=K$. 
\end{lemma}

\begin{proof}
As above, we condition on $X_i, Y_i$ never being simultaneously equal to $1$. Hence $X_i+Y_i$ is a Bernoulli random variable. Let us define $S = \sum (X_i + Y_i)$, which as a sum of independent Bernoulli variables has a binomial distribution.

For all $k$, and for $(a_i, b_i)$ satisfying $b_1 \geq b_2 \geq \cdots \geq b_k \geq 0$, $a_i + b_i = a_1+b_1+i-1$, we have:
$$\Pr[Z_{k+1} =  \varnothing, Z_k = (a_k, b_k), Z_{k-1} = (a_{k-1}, b_{k-1}), \ldots, Z_1 = (a_1, b_1)] = \frac{\Pr[S = a_k+b_k]}{2^{a_k+b_k}};$$
This is because given $a_k+b_k$, the total number of $1$'s should be $S = a_k+b_k$, and then the sequence $(a_1,b_1),\ldots,(a_k,b_k)$ uniquely determines which $1$ should be an $X_i$ and which one should be an $Y_i$.

Applying the formula above (with $k$ replaced  by $k+1$ and summing over the $b_{k}+1$ choices for $b_{k+1}$), we also get:
$$\Pr[Z_{k+2 }= \varnothing, Z_{k+1}  \neq \varnothing, Z_k = (a_k, b_k), Z_{k-1} = (a_{k-1}, b_{k-1}), \ldots, Z_1 = (a_1, b_1)] = \frac{b_k+1}{2^{a_k+b_k+1}} \Pr[S = a_k+b_k+1].$$

It follows that 
\begin{align*}
    \frac{\Pr[Z_{k+1} \neq \varnothing, Z_{k+2} =\varnothing \mid \mathcal{G}_k]}{\Pr[Z_{k+1} = \varnothing\mid \mathcal{G}_k]} 
    &= \frac{(B_k+1) \Pr[S=A_k+B_k+1]}{2\Pr[S=A_k+B_k]}
\end{align*}

Now since $B_{k+1} \leq B_k$, and $A_{k+1}+ B_{k+1}= A_k+B_k+1$, and the distribution of $S$ is log-concave, it follows that the ratio we just considered is non-increasing in $k$. Then, applying Theorem 1 in \cite{Ferguson}, it follows that the optimal stopping rule is the one that stops whenever 
$$  \frac{\Pr[Z_{k+1} \neq \varnothing, Z_{k+2} =\varnothing \mid Z_k]}{\Pr[Z_{k+1} = \varnothing\mid Z_k]} \leq 1.$$
In particular, the optimal stopping rule has the following form: Given the current pair of observations $A_k=a, B_k=b$, stop if and only if
$$\frac{(b+1) \Pr[S=a+b+1]}{2\Pr[S=a+b]}\leq 1.$$

Let us say that the stopping rule \textit{accepts} $(a, b)$ if this inequality is true. Due to the monotonicity we just demonstrated, the optimal stopping rule is \textit{monotone} in the sense that if we were ever to predict after observing $(a,b)$ that the stochastic process hits $\varnothing$ in the next stage, we would continue to predict that the stochastic process would hit $\varnothing$ if we were to observe $(a', b')$ after $(a, b)$. Hence we can analyze the probability of this stopping rule's success in the following way (quite similar to our analysis in the random-order variant, though we explain it in different words here):

The probability that the algorithm accepts at some point is exactly the probability that the algorithm would accept if presented with the last occurrence of a nonzero $X_i$, which is
$$\sum_{(a, b)\in A}\frac{1}{2^{b+1}} \Pr[S=a+b] $$
where $A$ is the set of $(a,b)$ values which the stopping rule accepts. This follows from the fact that 
$\frac{1}{2^{b+1}} \Pr[S=a+b]$ is exactly the probability that the configuration at the last occurrence of a nonzero $X_i$ is $(a,b)$; there should be $a+b$ total $1$s, and there should be exactly $b$ nonzero $Y$'s after the last nonzero $X_i$.

The probability that the stopping rule makes the wrong prediction at some point is
$$\sum_{(a, b)\in A} \frac{b+1}{2^{b+2}} \Pr[S=a+b+1]$$
since again by the monotonicity property, this is exactly the probability that the stopping rule would accept if presented with the penultimate nonzero $X_i$; the probability that this happens with a particular configuration $(a,b)$ is exactly the probability that there are $(b+1)$ 1's after this penultimate $X_{K-1}=1$, and exactly one of them is the last one, $X_K = 1$, which happens with probability $\frac{b+1}{2^{b+2}} \Pr[S=a+b+1]$. 

Hence, the stopping rule's success probability is
\begin{align*}
   &\sum_{(a, b)\in A}\frac{1}{2^{b+1}} \Pr[S=a+b] - \sum_{(a, b)\in A}\frac{b+1}{2^{b+2}} \Pr[S=a+b+1] \\ 
   &= \sum_{(a, b)\in A} \left(\frac{1}{2^{b+1}}-\frac{b+1}{2^{b+2}} \right) \Pr[S=a+b] +
   \sum_{(a, b)\in A}\frac{b+1}{2^{b+2}} (\Pr[S=a+b]-\Pr[S=a+b+1]) \\
   &\leq \sum_{(a,0)\in A} \frac{1}{4} \Pr[S=a] +\sum_{(a, b)} \frac{b+1}{2^{b+2}} |\Pr[S=a+b]-\Pr[S=a+b+1]|\\
   &\leq \frac{1}{4}+ \sum_{b}\frac{b+1}{2^{b+2}}\sum_a|\Pr[S=a+b]-\Pr[S=a+b+1]|\\
   &\leq \frac{1}{4}+ \sum_{b}\frac{b+1}{2^{b+2}}\left(2\max_c \Pr[S=c]\right)\\
   &\leq \frac{1}{4}+ 4\max_c \Pr[S=c]
\end{align*}
where the inequality $\sum_a|\Pr[S=a+b]-\Pr[S=a+b+1]|\leq 2\max_c \Pr[S=c])$ follows from telescoping, and noting that the sequence of probabilities is unimodal. Finally, $\max_c \Pr[S=c] \rightarrow 0$ as $n \rightarrow \infty$, since $S$ has a binomial distribution with expectation $\omega(1)$.
\end{proof}

Now consider a general algorithm for the single sample last success problem where the input is of large length $N$, and let us consider its behavior in the situation when there does not exist any $j$ for which $X_j = Y_j = 1$. Let us define $Q = \{(t, T) \mid t \in T \subseteq [N]\}$.

Suppose the algorithm were to observe $X_t = 1$ at time $t$, and $\sum_{i=1}^{t-1} (X_i + Y_i) = a$.
We claim that without loss of generality, the behavior of the algorithm does not depend on which of the $a$ past $1$'s are $X_i$'s or $Y_i$'s. This is because any algorithm $\cA$ can be replaced step by step by an algorithm $\cA'$ which, given any such a configuration, makes a prediction that $\cA$ would make if each of the $1$'s in the past were randomly assigned to be $X_i$ or $Y_i$ (a uniformly random one of the $2^a$ configurations). Since the choice of one of these configurations does not affect the distribution of the input in the future, and each of the $2^a$ configurations is equally likely to appear on the input, the expected performance of $\cA'$ is the same as that of $\cA$.

It follows that any algorithm for the problem can be thought of as function $f: Q \to [0,1]$, by thinking of the probability that the algorithm accepts the 1 it observes at position $t$ given the set of positions of all the  1s it has observed (amongst $Y_1, Y_2, \ldots, Y_N$ and $X_1, X_2, \ldots X_t$) is $T$ as $f((t, T))$, which we will write in short as $f(t, T)$.

Our goal now is to show that even though $f$ can be incredibly complicated, there is a large subset $S$ of $[N]$ on which $f$ actually looks like a stopping rule of the type we have considered before. 
Therefore, we prove the following lemma about functions from $Q$ to $[0,1]$ using the hypergraph Ramsey theorem\footnote{Unfortunately, this argument, relying as it does on the hypergraph Ramsey theorem does not give us particularly good bounds for the optimal probability of success for the single sample last success problem for $N = 100$ say.} (with a proof similar to Lemma 2.9 in \cite{obliviousOCRS}):

\begin{lemma}
Given any $\epsilon > 0$ and any natural number $n$, there is an $N$ so large that for any function $f: Q \to [0,1]$, there is a set $S \subset [N]$ of size $n$, such that for any $T_1, T_2 \subset S$ with $|T_1| = |T_2|$, we have $|f(t_{i,1}, T_1) - f(t_{i,2}, T_2)| \leq \epsilon$, where $t_{i, j}$ is the $i^{\text{th}}$ largest element in $T_j$.
\end{lemma}

\begin{proof}
We will actually prove the following slightly stronger statement (because it is easier to demonstrate by induction):

\begin{itemize}
    \item Given any $\epsilon > 0$ and any natural number $n, m$, there is an $N$ so large that for any function $f: Q \to [0,1]$, there is a set $S \subset [N]$ of size $n$, such that for any $T_1, T_2 \subset S$ with $|T_1| = |T_2|\leq m$, we have $|f(t_{i,1}, T_1) - f(t_{i,2}, T_2)| \leq \epsilon$, where $t_{i, j}$ is the $i^{\text{th}}$ largest element in $T_j$.
\end{itemize}

The proof is a simple induction on $m$. For $m = 1$, imagine coloring each number in $t \in [N]$ with the color $\lfloor\frac{f(t,\{t\})}{\epsilon}\rfloor$. The required claim is then just an immediate consequence of the pigeonhole principle.

Now imagine we have proven the statement for $m = k-1$, and we wish to prove the statement for $m = k$. 

By the hypergraph Ramsey theorem, we know that there exists $n_0$ so that any coloring of the complete $k$-hypergraph on $n_0$ vertices with $\left(\lfloor\frac{1}{\epsilon}\rfloor+1\right)^k$ colors has a monochromatic clique of size $n$.

By the induction hypothesis (applied with $n = n_0$), there exists $N$ so large that for any $f$, there is a subset $S_0 \subset [N]$ of size $n_0$ so that for any $T_1, T_2 \subset S$ and any $i$, with $|T_1| = |T_2|\leq k-1$, we have $|f(t_{i,1}, T_1) - f(t_{i,2}, T_2)| \leq \epsilon$. Therefore, we just need to find a further subset of $S_0$ of size $n$, on which $|f(t_{i,1}, T_1) - f(t_{i,2}, T_2)| \leq \epsilon$ is also true for $|T_1| = |T_2|=k$.

Consider the complete $k$-hypergraph with vertices as the number in the set $S_0$. Color the edge $T \subset S_0$ with the color $$\left(\lfloor\frac{f(t_1,T)}{\epsilon}\rfloor,\lfloor\frac{f(t_2,T)}{\epsilon}\rfloor,\ldots, \lfloor\frac{f(t_k, T)}{\epsilon}\rfloor\right)$$ where $t_1<t_2<\cdots <t_k$ are the elements of $T$. The result then follows immediately by Ramsey's theorem.
\end{proof}

What does this lemma tell us about $f$? Fix an $n$, and suppose we make sure that $X_i = 0$ with probability 1 for $i \in [N]\setminus S$, and in addition we make $\epsilon$ extremely small. Then $f$, up to an error $\epsilon$, is an algorithm for accepting the last 1 in $S$ with the property that when it sees $X_{s(t)} = 1$ and has to decide whether to accept it or not,  it does not look at the exact position of the 1s that it has seen, but only the number of 1s amongst $X_{s(1)}, X_{s(2)}, \ldots, X_{s(t)}$ and $Y_{s(1)}, Y_{s(2)}, \ldots, Y_{s(t-1)}$, and the number of 1s amongst $Y_{s(t+1)}, Y_{s(t+2)}, \ldots, Y_{s(n)}$, where $s(1) < s(2) <\cdots <s(n)$ is the set of numbers in $S$. But this is exactly the kind of stopping rule for which we have already demonstrated that it is impossible to do strictly better than $\frac{1}{4}$ for, by setting $X_i = 1$ with probability $\frac{1}{n^{2/3}}$ for $i \in S$. The desired result follows.

\paragraph{Acknowledgements.}
We would like to thank José Correa for enlightening discussions regarding the adversarial order two-sided Game of Googol, and the differences between their model in \cite{Independentsampling} and our own.

\printbibliography

@inproceedings{Independentsampling, author = {Correa, Jos\'{e} and Cristi, Andr\'{e}s and Feuilloley, Laurent and Oosterwijk, Tim and Tsigonias-Dimitriadis, Alexandros}, title = {The Secretary Problem with Independent Sampling}, year = {2021}, isbn = {9781611976465}, publisher = {Society for Industrial and Applied Mathematics}, address = {USA}, booktitle = {Proceedings of the Thirty-Second Annual ACM-SIAM Symposium on Discrete Algorithms}, pages = {2047–2058}, numpages = {12}, location = {Virtual Event, Virginia}, series = {SODA '21} }

@inbook{advice, author = {D\"{u}tting, Paul and Lattanzi, Silvio and Paes Leme, Renato and Vassilvitskii, Sergei}, title = {Secretaries with Advice}, year = {2021}, isbn = {9781450385541}, publisher = {Association for Computing Machinery}, address = {New York, NY, USA}, url = {https://doi.org/10.1145/3465456.3467623}, booktitle = {Proceedings of the 22nd ACM Conference on Economics and Computation}, pages = {409–429}, numpages = {21} }

@InProceedings{me1,
author="Nuti, Pranav",
editor="Aardal, Karen
and Sanit{\`a}, Laura",
title="The Secretary Problem with Distributions",
booktitle="Integer Programming and Combinatorial Optimization",
year="2022",
publisher="Springer International Publishing",
address="Cham",
pages="429--439",
isbn="978-3-031-06901-7"
}

@article{CS81,
 ISSN = {00018678},
 URL = {http://www.jstor.org/stable/1426783},
 author = {Gregory Campbell and Stephen M. Samuels},
 journal = {Advances in Applied Probability},
 number = {3},
 pages = {510--532},
 publisher = {Applied Probability Trust},
 title = {Choosing the Best of the Current Crop},
 urldate = {2022-07-12},
 volume = {13},
 year = {1981}
}

@article{Bruss2000,
author = {F. Thomas Bruss},
title = {{Sum the odds to one and stop}},
volume = {28},
journal = {The Annals of Probability},
number = {3},
publisher = {Institute of Mathematical Statistics},
pages = {1384 -- 1391},
year = {2000},
doi = {10.1214/aop/1019160340},
URL = {https://doi.org/10.1214/aop/1019160340}
}

@inbook{obliviousOCRS,
author = {Hu Fu and Pinyan Lu and Zhihao Gavin Tang and Abner Turkieltaub and Hongxun Wu and Jinzhao Wu and Qianfan Zhang},
title = {Oblivious Online Contention Resolution Schemes},
booktitle = {Symposium on Simplicity in Algorithms (SOSA)},
chapter = {},
pages = {268-278},
doi = {10.1137/1.9781611977066.20},
URL = {https://epubs.siam.org/doi/abs/10.1137/1.9781611977066.20},
eprint = {https://epubs.siam.org/doi/pdf/10.1137/1.9781611977066.20}
}

@inproceedings{Correaramsey, author = {Correa, Jos\'{e} and D\"{u}tting, Paul and Fischer, Felix and Schewior, Kevin}, title = {Prophet Inequalities for I.I.D. Random Variables from an Unknown Distribution}, year = {2019}, isbn = {9781450367929}, publisher = {Association for Computing Machinery}, address = {New York, NY, USA}, url = {https://doi.org/10.1145/3328526.3329627}, doi = {10.1145/3328526.3329627}, booktitle = {Proceedings of the 2019 ACM Conference on Economics and Computation}, pages = {3–17}, numpages = {15}, location = {Phoenix, AZ, USA}, series = {EC '19} }

@article{Ferguson,
author = {Thomas S. Ferguson},
journal = {Mathematica Applicanda},
language = {eng},
number = {1},
pages = {null},
title = {The Sum-the-Odds Theorem with Application to a Stopping Game of Sakaguchi},
url = {http://eudml.org/doc/292688},
volume = {44},
year = {2016},
}

@inproceedings{CDFFLLP22,
  author    = {Constantine Caramanis and
               Paul D{\"{u}}tting and
               Matthew Faw and
               Federico Fusco and
               Philip Lazos and
               Stefano Leonardi and
               Orestis Papadigenopoulos and
               Emmanouil Pountourakis and
               Rebecca Reiffenh{\"{a}}user},
  editor    = {Joseph (Seffi) Naor and
               Niv Buchbinder},
  title     = {Single-Sample Prophet Inequalities via Greedy-Ordered Selection},
  booktitle = {Proceedings of the 2022 {ACM-SIAM} Symposium on Discrete Algorithms,
               {SODA} 2022, Virtual Conference / Alexandria, VA, USA, January 9 -
               12, 2022},
  pages     = {1298--1325},
  publisher = {{SIAM}},
  year      = {2022},
  url       = {https://doi.org/10.1137/1.9781611977073.54},
  doi       = {10.1137/1.9781611977073.54},
  bibsource = {dblp computer science bibliography, https://dblp.org}
}

@article{DRY15,
  author    = {Peerapong Dhangwatnotai and
               Tim Roughgarden and
               Qiqi Yan},
  title     = {Revenue maximization with a single sample},
  journal   = {Games Econ. Behav.},
  volume    = {91},
  pages     = {318--333},
  year      = {2015},
  url       = {https://doi.org/10.1016/j.geb.2014.03.011},
  doi       = {10.1016/j.geb.2014.03.011},
  bibsource = {dblp computer science bibliography, https://dblp.org}
}

@inproceedings{DFLLR21a,
  author    = {Paul D{\"{u}}tting and
               Federico Fusco and
               Philip Lazos and
               Stefano Leonardi and
               Rebecca Reiffenh{\"{a}}user},
  editor    = {Samir Khuller and
               Virginia Vassilevska Williams},
  title     = {Efficient two-sided markets with limited information},
  booktitle = {{STOC} '21: 53rd Annual {ACM} {SIGACT} Symposium on Theory of Computing,
               Virtual Event, Italy, June 21-25, 2021},
  pages     = {1452--1465},
  publisher = {{ACM}},
  year      = {2021},
  url       = {https://doi.org/10.1145/3406325.3451076},
  doi       = {10.1145/3406325.3451076}
  }

@article{DFLLR21b,
  author    = {Paul D{\"{u}}tting and
               Federico Fusco and
               Philip Lazos and
               Stefano Leonardi and
               Rebecca Reiffenh{\"{a}}user},
  title     = {Prophet Inequalities for Matching with a Single Sample},
  journal   = {CoRR},
  volume    = {abs/2104.02050},
  year      = {2021},
  url       = {https://arxiv.org/abs/2104.02050},
  eprinttype = {arXiv},
  eprint    = {2104.02050},
  bibsource = {dblp computer science bibliography, https://dblp.org}
}

@InProceedings{Superstars, title = {Prophets, Secretaries, and Maximizing the Probability of Choosing the Best}, author = {Esfandiari, Hossein and Hajiaghayi, MohammadTaghi and Lucier, Brendan and Mitzenmacher, Michael}, booktitle = {Proceedings of the Twenty Third International Conference on Artificial Intelligence and Statistics}, pages = {3717--3727}, year = {2020}, editor = {Silvia Chiappa and Roberto Calandra}, volume = {108}, series = {Proceedings of Machine Learning Research}, address = {Online}, month = {8}, publisher = {PMLR}, url = {http://proceedings.mlr.press/v108/esfandiari20a.html}  }

@InProceedings{rubinstein,
  author =	{Aviad Rubinstein and Jack Z. Wang and S. Matthew Weinberg},
  title =	{{Optimal Single-Choice Prophet Inequalities from Samples}},
  booktitle =	{11th Innovations in Theoretical Computer Science Conference (ITCS 2020)},
  pages =	{60:1--60:10},
  series =	{Leibniz International Proceedings in Informatics (LIPIcs)},
  ISBN =	{978-3-95977-134-4},
  ISSN =	{1868-8969},
  year =	{2020},
  volume =	{151},
  editor =	{Thomas Vidick},
  publisher =	{Schloss Dagstuhl--Leibniz-Zentrum fuer Informatik},
  address =	{Dagstuhl, Germany},
  URL =		{https://drops.dagstuhl.de/opus/volltexte/2020/11745},
  URN =		{urn:nbn:de:0030-drops-117452},
  doi =		{10.4230/LIPIcs.ITCS.2020.60}
}

@inproceedings{googol,
  author={José R. Correa and Andrés Cristi and Boris Epstein and José A. Soto},
  title={The Two-Sided Game of Googol and Sample-Based Prophet Inequalities},
  year={2020},
  cdate={1577836800000},
  pages={2066-2081},
  url={https://doi.org/10.1137/1.9781611975994.127},
  booktitle={SODA},
}

@article{sampledriven,
  author    = {Jos{\'{e}} R. Correa and
               Andr{\'{e}}s Cristi and
               Boris Epstein and
               Jos{\'{e}} A. Soto},
  title     = {Sample-driven optimal stopping: From the secretary problem to the
               i.i.d. prophet inequality},
  journal   = {CoRR},
  volume    = {abs/2011.06516},
  year      = {2020},
  url       = {https://arxiv.org/abs/2011.06516},
  archivePrefix = {arXiv},
  eprint    = {2011.06516},
  bibsource = {dblp computer science bibliography, https://dblp.org}
}

\end{document}